\documentclass[conference]{IEEEtran}
\IEEEoverridecommandlockouts
\usepackage{cite}
\usepackage{amsmath,amssymb,amsfonts}
\usepackage{graphicx}
\usepackage{textcomp}
\usepackage{xcolor}
\usepackage{amsthm}
\usepackage{algorithm}
\usepackage{algorithmicx}  
\usepackage{algpseudocode}
\usepackage[caption=false]{subfig}

\def\BibTeX{{\rm B\kern-.05em{\sc i\kern-.025em b}\kern-.08em
    T\kern-.1667em\lower.7ex\hbox{E}\kern-.125emX}}

\begin{document}

\title{A System of Monitoring and Analyzing Human Indoor Mobility and Air Quality
}

\author{
\IEEEauthorblockN{Kyle K. Qin}
\IEEEauthorblockA{\textit{School of Computing Technologies} \\
\textit{RMIT University}\\
kyle.qin@hotmail.com}
\and
\IEEEauthorblockN{Mohammad S. Rahaman}
\IEEEauthorblockA{\textit{School of Computing Technologies} \\
\textit{RMIT University}\\
saiedur.rahaman@rmit.edu.au}
\and
\IEEEauthorblockN{Yongli Ren}
\IEEEauthorblockA{\textit{School of Computing Technologies} \\
\textit{RMIT University}\\
yongli.ren@rmit.edu.au}
\and
\IEEEauthorblockN{Chi-Tsun Cheng}
\IEEEauthorblockA{\textit{School of Engineering} \\
\textit{RMIT University}\\
ben.cheng@rmit.edu.au}
\and
\IEEEauthorblockN{Ivan Cole}
\IEEEauthorblockA{\textit{School of Engineering} \\
\textit{RMIT University}\\
ivan.cole@rmit.edu.au}
\and
\IEEEauthorblockN{Flora D. Salim}
\IEEEauthorblockA{\textit{School of Computer Science} \\
\textit{University of New South Wales}\\
flora.salim@unsw.edu.au}
}

\maketitle

\begin{abstract}
Human movements in the workspace usually have non-negligible relations with air quality parameters (e.g., CO$_2$, PM2.5, and PM10). We establish a system to monitor indoor human mobility with air quality and assess the interrelationship between these two types of time series data. More specifically, a sensor network was designed in indoor environments to observe air quality parameters continuously. Simultaneously, another sensing module detected participants' movements around the study areas. In this module, modern data analysis and machine learning techniques have been applied to reconstruct the trajectories of participants with relevant sensor information. Finally, a further study revealed the correlation between human indoor mobility patterns and indoor air quality parameters. Our experimental results demonstrate that human movements in different environments can significantly impact air quality during busy hours. With the results, we propose recommendations for future studies.
\end{abstract}


\section{Introduction}
In recent years, many studies focused on the association between human mobility and the changes in air quality, together with the implication of their relation to COVID-19 or other airborne disease transmissions \cite{adam2021air, zhu2020mediating, li2020changes, archer2020changes}. Governments introduced various measures to limit people's mobility to protect public health, such as restricted road and air travel. The curtailment of personal travel and various economic activities contributes to the reductions in the atmospheric levels of outdoor air pollutants in many countries, such as particulate matter (PM), nitrogen dioxide (NO$_2$), carbon monoxide (CO), sulfur dioxide (SO$_2$), and volatile organic compounds (VOCs) \cite{adam2021air}. Air quality is closely related to the risk of COVID-19 infection because the body's immunity could be affected by aerosols that carry the virus \cite{bianconi2020particulate}. The study on air quality patterns can help to understand the association of human mobility with virus infection \cite{zhu2020mediating}. Compared with the outdoor environments, the emissions of PM and VOCs from intensive indoor activities carried out by the occupants could potentially expose residents to more harm \cite{adam2021air, li2022field}. Nowadays, indoor air quality (IAQ) has become an increasingly important concern for human health, which is often linked to allergies and airway infections \cite{jones1999indoor, sundell2004history, pillarisetti2022indoor}.

Heretofore, various systems based on the Internet of Things (IoT) techniques have been established for indoor air quality monitoring purposes \cite{jiang2011maqs, benammar2017real, alhakbani2013zone, xiang2013hybrid, kim2014issaq, mumtaz2021internet}. IAQ monitoring is challenging because indoor air pollutants concentration and human motion patterns vary spatially and temporally within the buildings. In the previous studies, researchers primarily built IAQ monitoring systems with stationary air quality monitoring sensors, which can provide accurate readings with effective periodic calibrations \cite{kim2014issaq, benammar2017real}. Modern IAQ monitoring systems typically integrate onsite air quality detectors which send sensor data to an interactive platform or cloud via nearby gateways \cite{alhmiedat2017low, zhou2020internet, evagelopoulos2021smart, li2021air, mumtaz2021internet, liu2021multi}. However, stationary sensors are usually more expensive for large-scale and fine-grained deployment, and the maintenance cost is relatively high and lacks certain flexibility of deployment \cite{xiang2013hybrid}. More recent works rely on mobile sensing networks for IAQ monitoring, requiring mobile communication modules and measurement instruments \cite{jiang2011maqs, xiang2013hybrid, alhakbani2013zone, maag2018w}. Mobile sensors are more susceptible to drift error (inaccurate reading) than stationary sensors due to difficulties in frequently calibrating the sensors, mainly when non-specialists use them.

\begin{figure*}[ht]
\centering
  \includegraphics[width=0.8\textwidth]{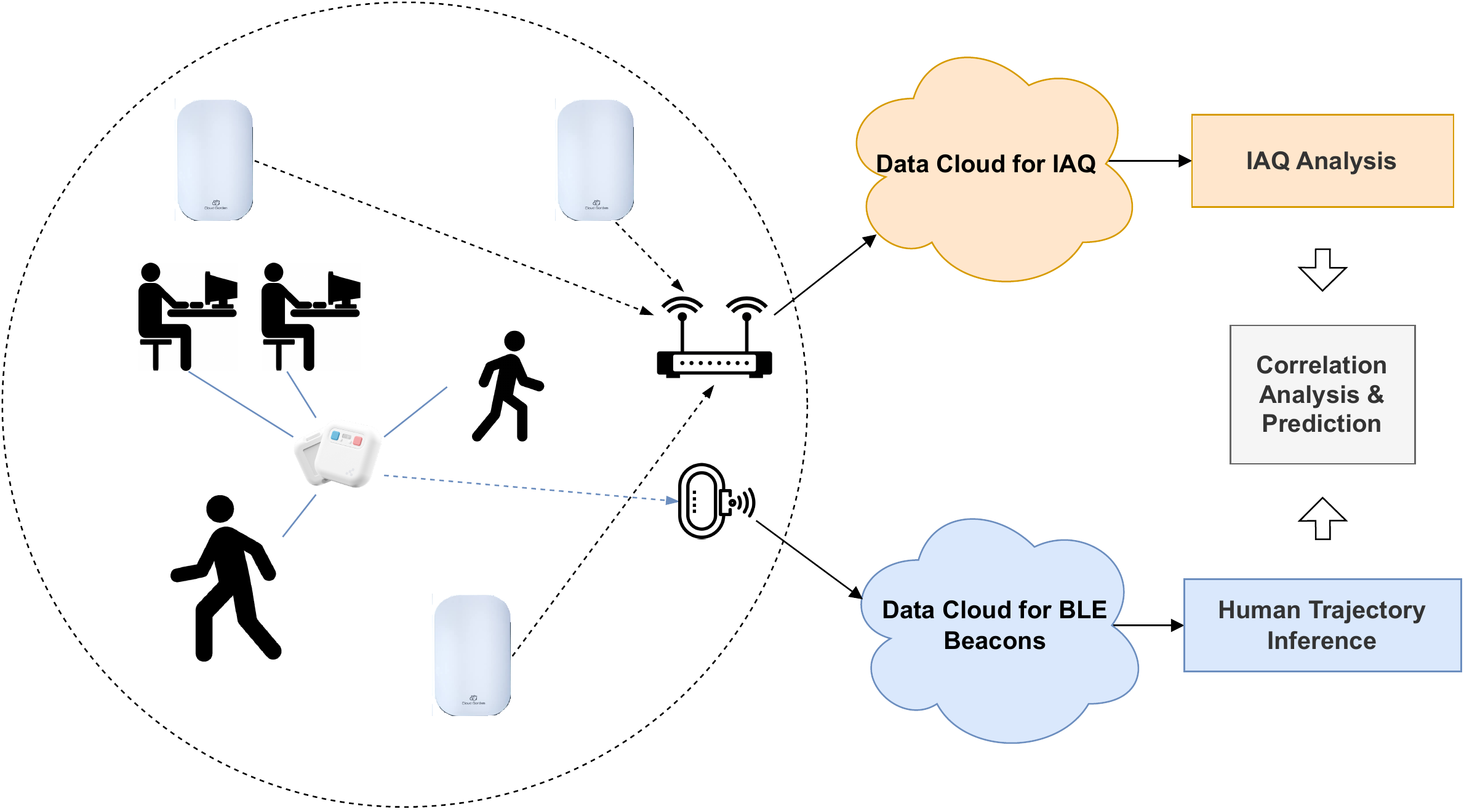}
  \caption{The overview of the integrated system for monitoring and analyzing IAQ with indoor human mobility.}
  \label{fig:Indoor-movement-IAQ}
\end{figure*}

The research on indoor localization aims to infer human positions in indoor environments. Modern localization technologies can be classified into Satellite‑Based Navigation \cite{xu2015new}, Magnetic Based Navigation \cite{xie2014maloc}, Optical Based Navigation \cite{raharijaona2017local}, Sound‑Based Technologies \cite{sertatil2012novel}, and Radio Frequency (RF) Based Navigation \cite{liu2014survey}. RF-based systems are adopted mostly for indoor localization, which can cover a broader area with low-cost hardware required. They often perform better than optical-based and ultrasonic-based navigations, as RF waves can penetrate materials such as walls and human bodies. RF-based navigation systems are commonly built with WiFi \cite{le2012wifi,ren2017d}, Bluetooth \cite{altini2010bluetooth,zafari2019survey}, Zigbee \cite{bianchi2018rssi}, and Radio Frequency Identification (RFID) \cite{tesoriero2010improving}. The popular localization detection methods include proximity-based techniques \cite{kupper2005location}, dead reckoning \cite{bao2014novel}, triangulation \cite{lin2013accurate}, and fingerprinting \cite{alkasi2013experimental,zhuang2016smartphone}. In addition, some works can also utilize the RSSI of sensors to predict the occupancy patterns of areas in indoor environments \cite{rahaman2019occuspace,zhang2022review}, which is significant for energy efficiency in the buildings.

The existing literature indicated that the spatiotemporal changes in air quality and individual movement patterns play an essential role in health risk assessment \cite{park2017individual}. Human activities can significantly affect indoor airflow patterns as human-induced movements quickly lead to the transport of contaminants \cite{bhattacharya2021three}. The published works examine the airflow fields around the moving body. They tend to concentrate on the tentative pattern of airflow characteristics with a moving human subject in an experimental environment with limited behaviors. In contrast, we intend to monitor and analyze the patterns of indoor air quality and real-time human movements in different workplaces. In this study, we designed an integrated system to monitor and analyze the IAQ and human mobility simultaneously in different sites, including a large work office and a factory workshop. Two surveillance modules were set up for data collection of the IAQ and human trajectories. Several stationary sensors were chosen for deployment in the study areas to gain a long-term observation of IAQ. Meanwhile, an anonymous mode is applied to detect participants' positions via Bluetooth Low Energy (BLE) beacon technology parallel to the running of IAQ sensors. Data collected from both modules were transmitted to the corresponding cloud platforms via nearby gateways in real time. In the analysis stage, we first infer the trajectories of anonymous participants by the fingerprinting approaches, which require the attendees to carry BLE beacons in monitoring spaces during working hours. In the following, relationship analytics were carried out on human trajectories and indoor air quality patterns.

Overall, the contributions of this work are summarized as follows:
\begin{itemize}
\item We designed a new integrated system based on two-type sensor networks to monitor and analyze the patterns of IAQ and human indoor movements. The system enables the inference of indoor human trajectories on anonymous participants during the monitoring stage.
\item The study's observation across multiple real-world workplaces, including a large work office and a factory workshop.
\item Intensive analysis was conducted on the flow patterns of the IAQ with the observed and inferred human displacement during working hours. The experimental results provide insights into the correlation between IAQ and indoor human mobility. A learning model has been established for analyzing the changes in air quality based on the shift characteristics of the participants.
\end{itemize}

The remainder of the paper is organized as follows. Section \ref{sec:system-overview} provides a high-level system overview. Section \ref{sec:experimental-results} describes the details of the experiments and the analysis results. The conclusion is given in Section \ref{sec:conclude}.

\section{The System Overview}
\label{sec:system-overview}

In the beginning, two monitoring modules are introduced to collect raw data for IAQ analysis and to track indoor participants in a work office and a factory workshop. In this stage, a specialized sensor network was set up for data collection, continuously detecting indoor air conditions. Meanwhile, another monitoring module is responsible for scanning the signals of BLE beacons carried by participants for movement inference. During the data collection, the IAQ data and BLE beacon signals will be transferred to their data clouds, respectively. After that, we present a method for constructing indoor human trajectories of the participants. Based on the patterns of IAQ and human movements found in the first analytical stage, we further develop an approach for correlation analysis between two types of observed patterns. Fig. \ref{fig:Indoor-movement-IAQ} gives an illustration of the whole system in detail.

\subsection{Sensor Network of Indoor Air Quality}
\begin{figure}[ht]
  \centering
    \subfloat[Air quality sensor
    \label{subfig:indoor_air_quality_sensor}]{
    \includegraphics[width=0.16\linewidth]{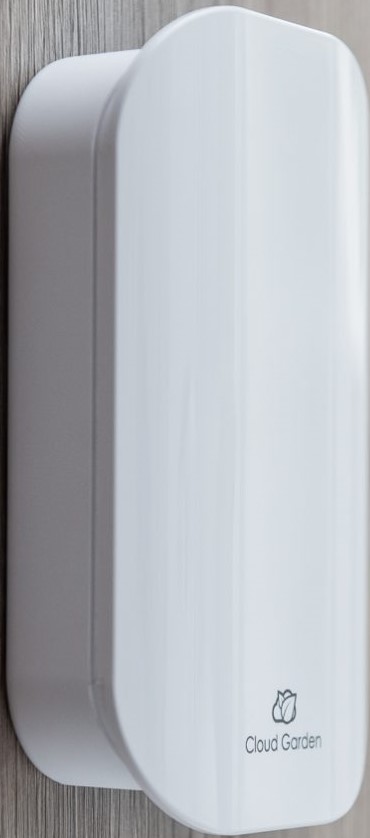}
    } 
    \hspace{0.3cm}
    \subfloat[Portal light
    \label{subfig:portal_light}]{
    \includegraphics[width=0.24\linewidth]{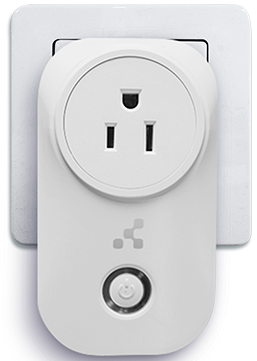}
    }
    \subfloat[BLE tags
    \label{subfig:tag}]{
    \includegraphics[width=0.25\linewidth]{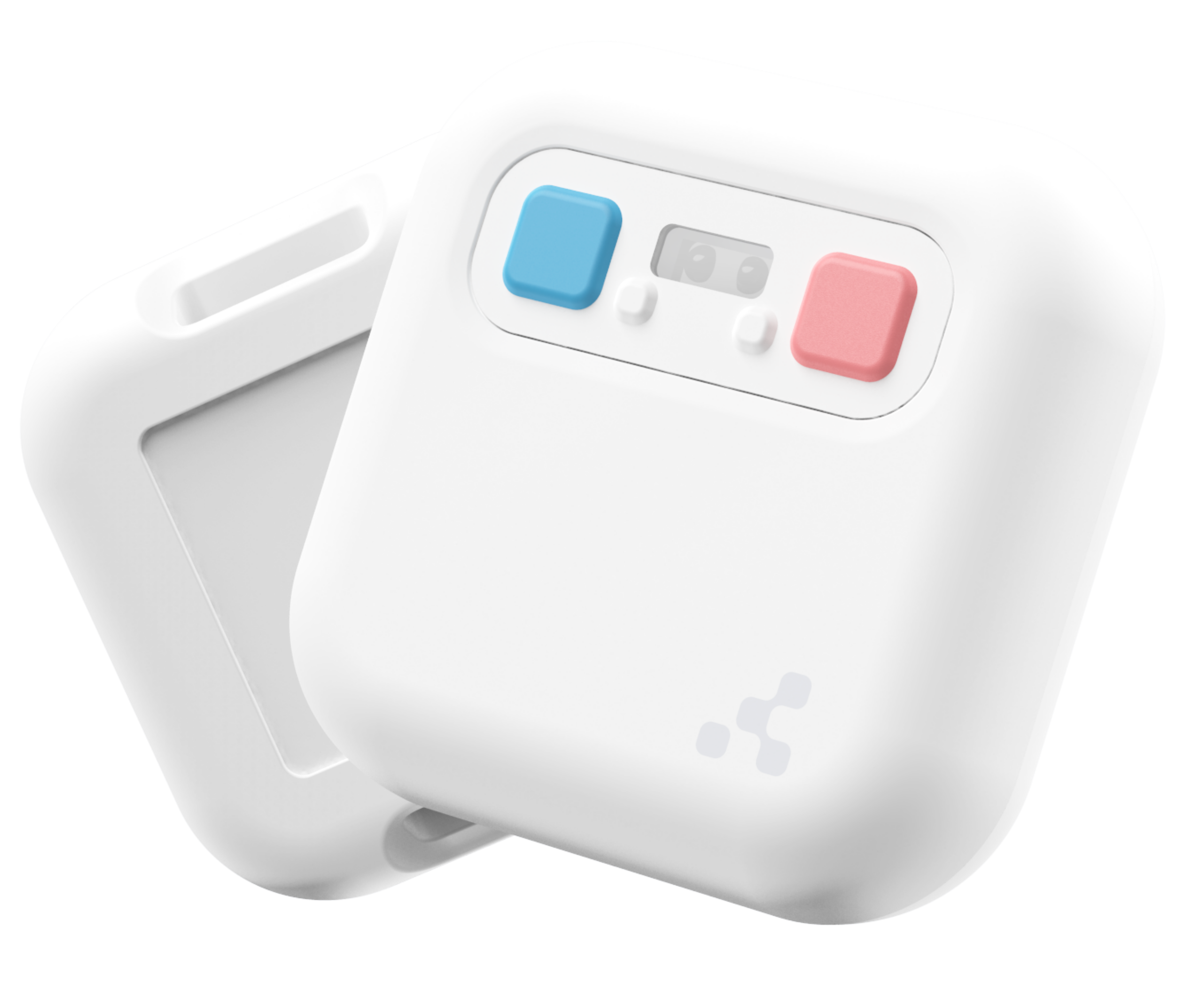}
    }
  \caption{Indoor climate sensors and tracking devices of human indoor mobility.}
  \label{fig:sensors}
\end{figure}

\begin{figure}[ht]
  \centering
    \subfloat[Work offices
    \label{subfig:memko_deploy}]{
    \includegraphics[width=0.83\linewidth]{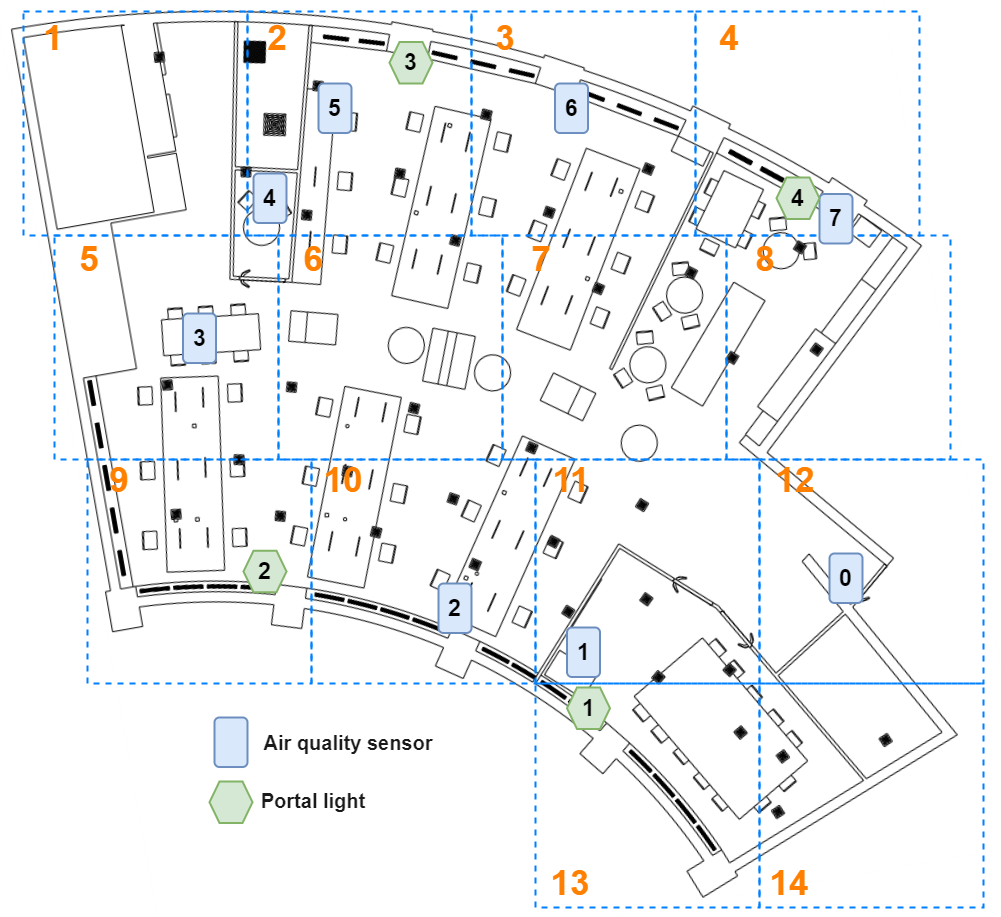}
    } \hspace{0.3cm}
    \subfloat[Factory workshop
    \label{subfig:ST_deploy}]{
    \includegraphics[width=0.8\linewidth]{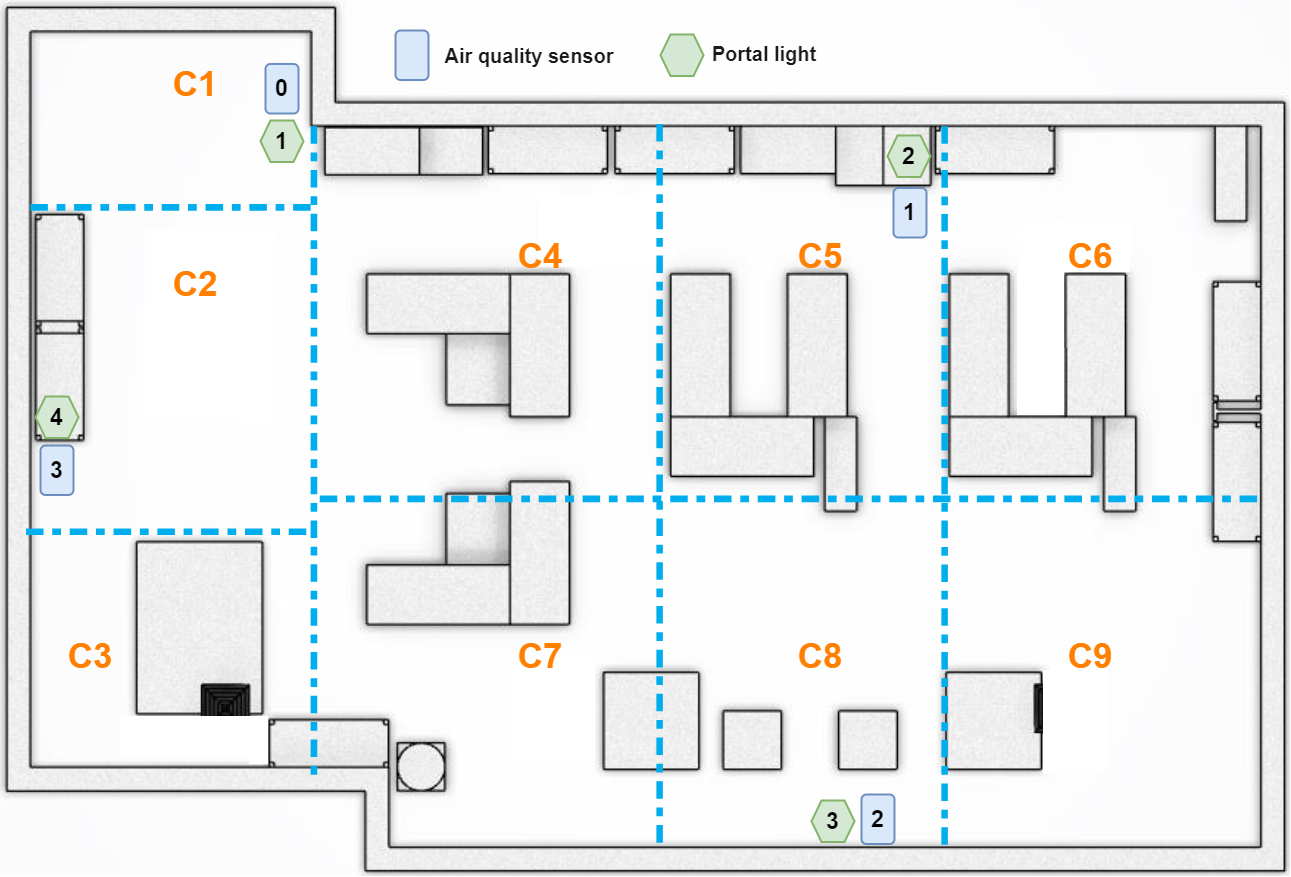}
    }
  \caption{The study areas and the deployment of indoor climate sensors and portal lights.}
  \label{fig:sensors_deploy}
\end{figure}

Firstly, all the hardware devices were verified to ensure that each was adequately prepared and ready for deployment. Then, a sensor network was designed and established to collect data on indoor climate conditions: PM, VOC, temperature, and CO$_2$. As Fig. \ref{subfig:indoor_air_quality_sensor} shows, air quality sensors are energy-efficient devices\footnote{https://cloudgarden.nl/en/product/climate-sensor/} that can monitor the nearby air conditions in houses and upload the data to the cloud via corresponding gateways. Fig. \ref{fig:sensors_deploy} demonstrates the deployment details of air quality sensors (in blue color) that are required for installation in different study areas, such as 7 of them in Memko offices and four devices in the Sutton Tools (ST) workshop.

\subsection{Tracking Network of Human Indoor Mobility}
In Fig. \ref{fig:sensors}, the Portal light\footnote{https://store.kontakt.io/product/portal-light-ac-us-plug-b/} is a BLE gateway that scans nearby BLE devices and sends received data such as Received Signal Strength Indicator (RSSI) to the cloud over its WiFi network connection. BLE beacon or tag\footnote{https://store.kontakt.io/product/asset-tag-2/} is an advanced BLE device for asset tracking and management. In our experiments, when anonymous participants carry these tags, we can infer the workplace location where a participant resides at a specific time. Therefore, participants must have the tags while staying at the observation workplace – the tag can be put in their pocket or attached to a lanyard. Specifically, tag signals will only be collected within the study areas that vary at different sites. No data collection will occur when the participants leave the monitoring sites. Participants can take the tags anywhere they want, and the system will only identify the tags once they enter the monitoring zones. The study is anonymous, and we aim to investigate whether human movements influence indoor air quality. In the end, we deployed a different number of Portal lights (in green color) at Memko offices and Sutton Tools to cover all the study areas, as shown in Fig. \ref{subfig:memko_deploy} and \ref{subfig:ST_deploy}.

\subsection{Human Trajectory Inference}
We implement fingerprinting techniques to infer the positions of anonymous participants in real-world workspaces. Fingerprinting technologies are famous for building indoor localization systems \cite{ni2010fingerprint, luo2017smartphone}. In a typical fingerprinting indoor localization system, $m$ signal readers are set up to scan nearby tags for tracking objects. The fingerprinting method utilize the received signal strength (RSS) information of observed tags by multiple RSSI readers to locate an unknown tag according to the similarity of signal patterns.

\subsubsection{Problem Description}
Let $Z = \{C_1, C_2, C_3, ..., C_n\}$ be the set of labels for zones created by dividing a study workplace into $n$ smaller grids with different sizes. A grid is considered occupied when a participant carrying a Bluetooth tag stays in the grid. The RSSI is recorded by nearby gateways (Portal lights) over time, which will be used as the fingerprints of tags in a specific grid. The RSSI of a tag captured by different gateways can be transferred to a feature vector for predicting the visited or occupied zone of a person with the tag. The goal is to build a classifier based on the observed RSSI features and corresponding zone labels to predict unknown human positions.

\begin{figure*}[ht!]
  \centering
    \subfloat[Grids prediction for carried tags.
    \label{subfig:grids_carry_tags_fingerprint_res}]{
    \includegraphics[width=0.3\textwidth]{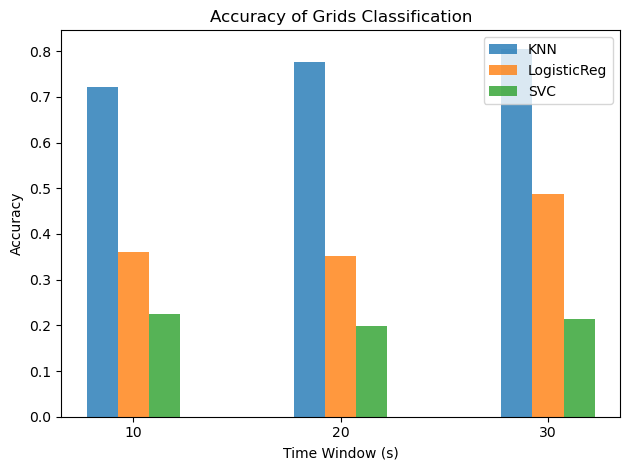}
    }
    \subfloat[Grids prediction for stationary tags. 
    \label{subfig:grids_fix_tags_fingerprint_res}]{
    \includegraphics[width=0.3\textwidth]{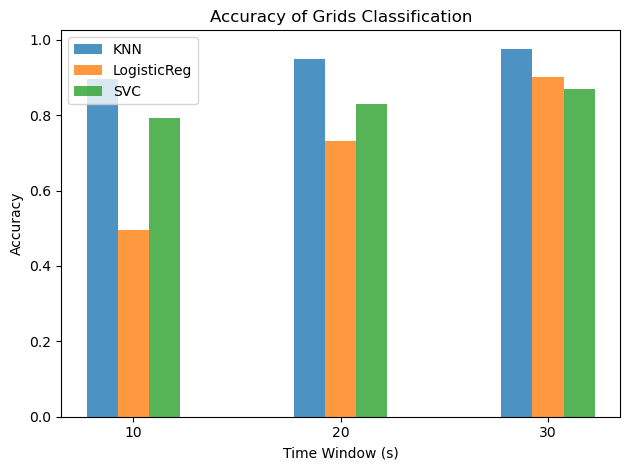}
    }
    \subfloat[Grids prediction with mixing data of carried and stationary tags.
    \label{subfig:grids_mix_fix_carry_fingerprint_res}]{
    \includegraphics[width=0.3\textwidth]{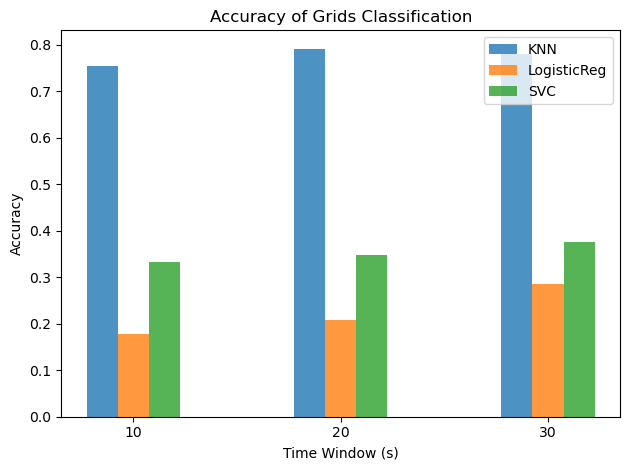}
    }    
  \caption{Position prediction for BLE tags.}
  \label{fig:grids_pred_res}
\end{figure*}

\subsubsection{Location Prediction}
In the stage of location prediction, raw RSSI values recorded from $m$ gateways are combined by the ID of Bluetooth tags. The collection of RSSI data of a tag is continuous over a defined period. Because not all the gateways receive the signal of a tag at the same timestamp, we applied time segmentation on the records of each tag by aggregating RSSI values in a short time window ($\Delta t$). Here, statistical features, including mean, first quartile, third quartile, ninth quartile, and median, are computed from the RSSI records in each time window (bounded by $\Delta t$). Simultaneously, the labels of occupied grids of participants are also segmented in the same time window. Therefore, the constructed features and corresponding grid labels are used to train classification models and predict locations for anonymous participants. 

To obtain the training dataset, we collected the fingerprints of tags in each grid for about $15-20$ minutes. However, there are some areas where it is not feasible to collect data for a long time (e.g., a field of consoles or wards of hospitals). To cover all the areas for study, we deployed stationary tags on the edge of physical assets in the workspace, such as tables, kitchens, and consoles, to imitate users with static behaviors. The data of fixed tags often last for $1-2$ days in each study site, which will be combined with those of carried tags by human testers for training models to predict all considered zones. We train a set of classifiers, including K-Nearest Neighbor (KNN), Support Vector Machine (SVM), and Logistic Regression. The best classifier is then used to infer the positions of anonymous participants at a given time.

\section{Experimental Results}
\label{sec:experimental-results}

\subsection{Outcome of Human Trajectory Construction}
We randomly separate the dataset, manually collected by our testers, into the training and testing sets to train the models. The evaluation metric is prediction accuracy which measures the ratio of correctly classified instances to the total number of samples. To have a complete evaluation, different case studies have been considered to verify the accuracy of tag location prediction. The details of the three tests are given as follows:
\begin{itemize}
\item In the first scenario, we carried three tags in each grid for around $15-20$ minutes in different sites. Using the fingerprints of those tags in grids for evaluation: $80\%$ for training and $20\%$ for testing the prediction of grid ID when the RSSI values of a carried tag are given. As Fig. \ref{subfig:grids_carry_tags_fingerprint_res} shows, the optimal model KNN can give an accuracy of 0.7 to 0.8 for time segmentations with 10 seconds, 20 seconds, and 30 seconds.
\item In the second case, we placed $3-5$ tags (mounted on tables) in each grid for $1-2$ days. The fingerprints of stationary tags were then used for training and testing: $80\%$ for training and $20\%$ for testing the prediction of grid ID when a fixed tag is given. Fig. \ref{subfig:grids_fix_tags_fingerprint_res} shows that with KNN, an accuracy higher than 0.9 can be obtained.
\item In the last experiment, we combine the fingerprints of stationary tags in the second case, the data of carried tags in the first test ($80\%$) for training, and $20\%$ data of taken tags are used for testing the prediction of grid ID. Noticeably, for a good balance between using the data of stationary and carried tags by testers for training, we limit the duration of data collection on stationary tags to around 20 minutes.
\end{itemize}

The tags carried by our human testers are often located at the center of each grid with standing and slightly walking behaviors. Using the fingerprints of stationary tags to predict a tag carried by people could allow us to cover all the grids in different sites and imitate the behaviors of people who remain motionless, like working at a station, for some time. The accuracy of the 20s time window is stable (staying at about 0.78) by KNN when comparing the prediction results between the first and third study cases. Therefore, we select the trained KNN in the third case as the final prediction model to infer the position of anonymous participants.

\subsection{Statistical Analysis}
We divided the working areas of Memko into 14 grids and 14 anonymous participants involved in this study. Staff participation days range from 4 to 8 on this site. In contrast, nine grids and 2 participants were in the Sutton Tools workshop, and the study lasted 24 days. It must be mentioned that not all the staff working on the sites can be monitored, as participation is voluntary. However, it is still valuable that we can deliver insights from the movement patterns of partial staff and their impact on indoor air quality. In the former step, we inferred participants' positions via fingerprinting approach. Then, we analyze occupancy patterns, visit frequency, and the correlation between human mobility and indoor air quality in the following subsections.

\subsubsection{Occupancy Rate}
\begin{figure}[t]
  \centering
    \subfloat[Memko - occupancy rate
    \label{subfig:mem_grid_occupy_portion}]{
    \includegraphics[width=0.45\linewidth]{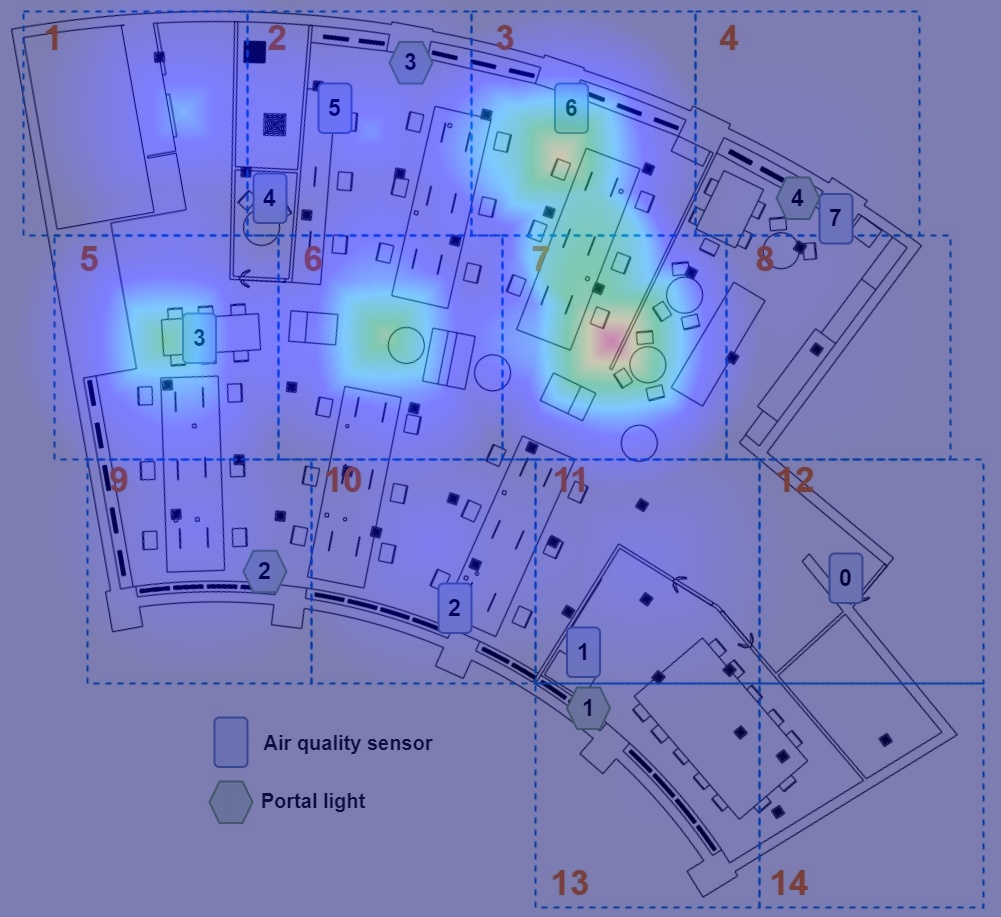}
    }
    \subfloat[Memko - visit frequency 
    \label{subfig:mem_grid_visit_freq}]{
    \includegraphics[width=0.45\linewidth]{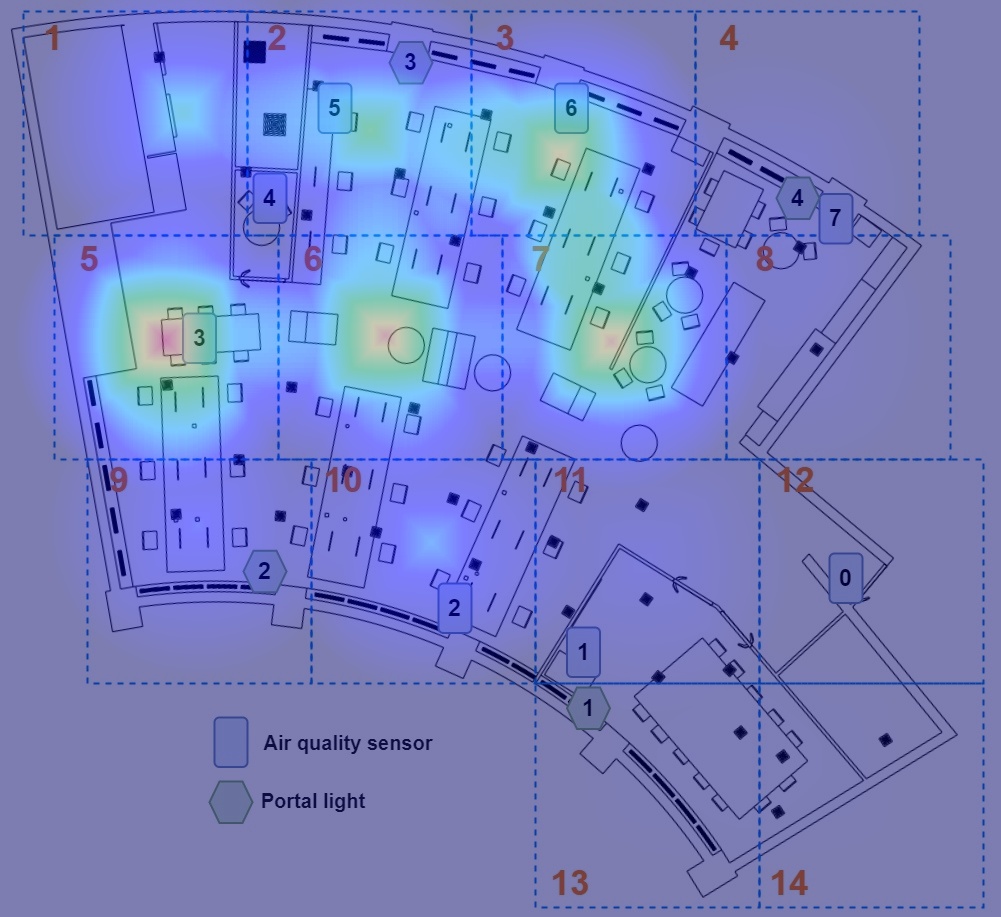}
    } \\
    \subfloat[Sutton Tools - occupancy rate
    \label{subfig:st_grid_occupy_portion}]{
    \includegraphics[width=0.45\linewidth]{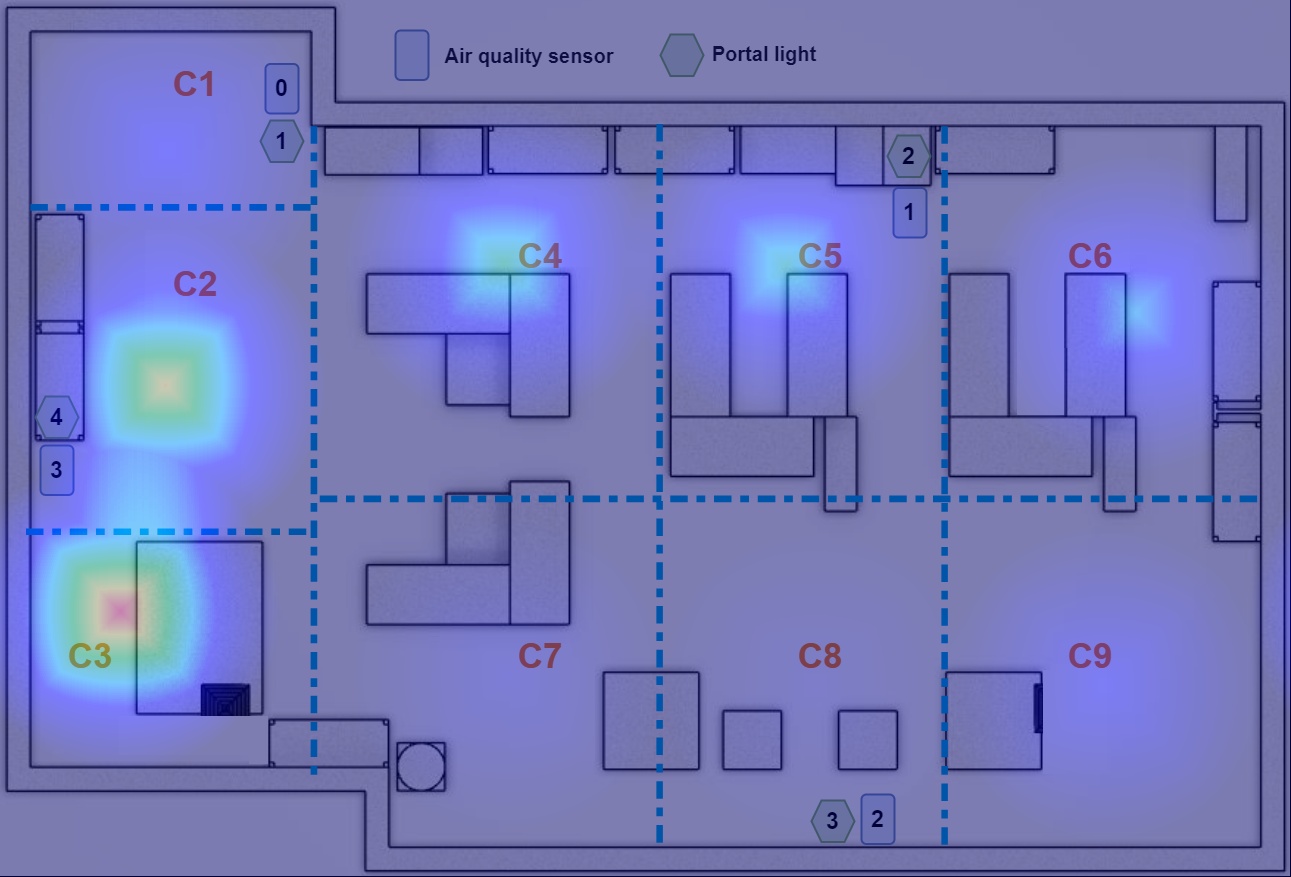}
    }
    \subfloat[Sutton Tools - visit frequency 
    \label{subfig:st_grid_visit_freq}]{
    \includegraphics[width=0.45\linewidth]{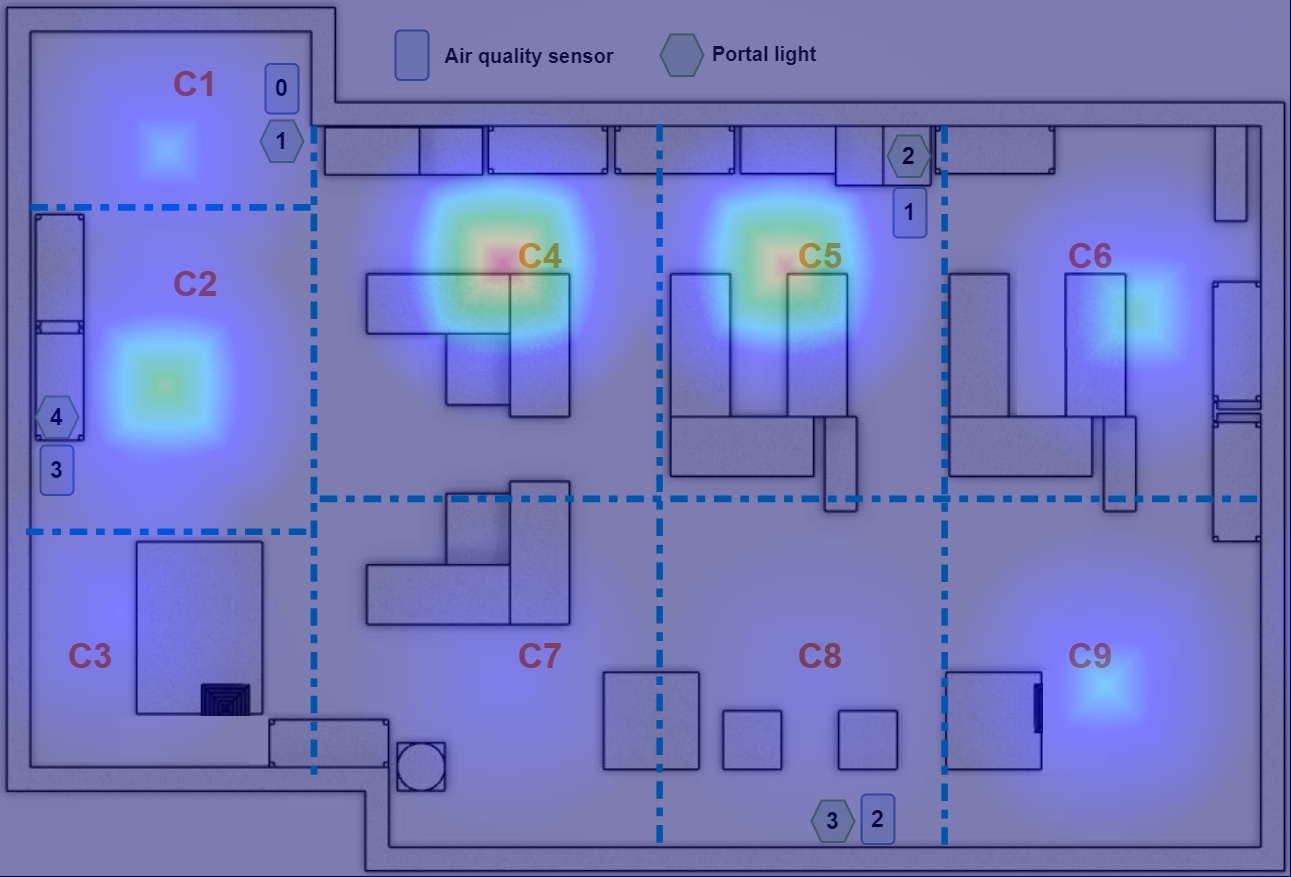}
    }
  \caption{Heatmap of occupancy rate and visit frequency by grids on two study areas.}
  \label{fig:occupancy_rate_visit_freq}
\end{figure}

\begin{figure}[t]
  \centering
    \subfloat[Memko - occupancy rate
    \label{subfig:memko-grid_occupy_portion_bar}]{
    \includegraphics[width=0.45\linewidth]{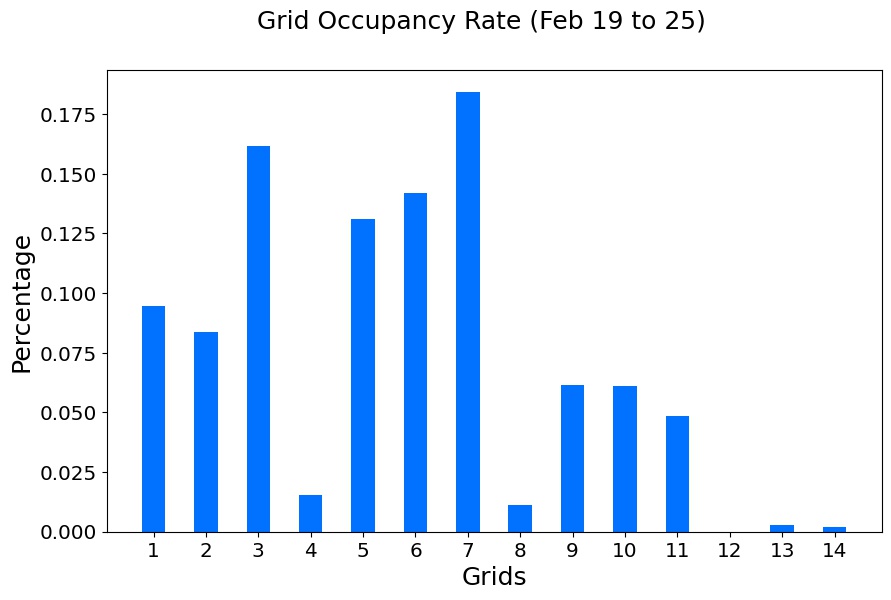}
    }
    \subfloat[Memko - visit frequency 
    \label{subfig:memko-grid_visiting_freq_bar}]{
    \includegraphics[width=0.45\linewidth]{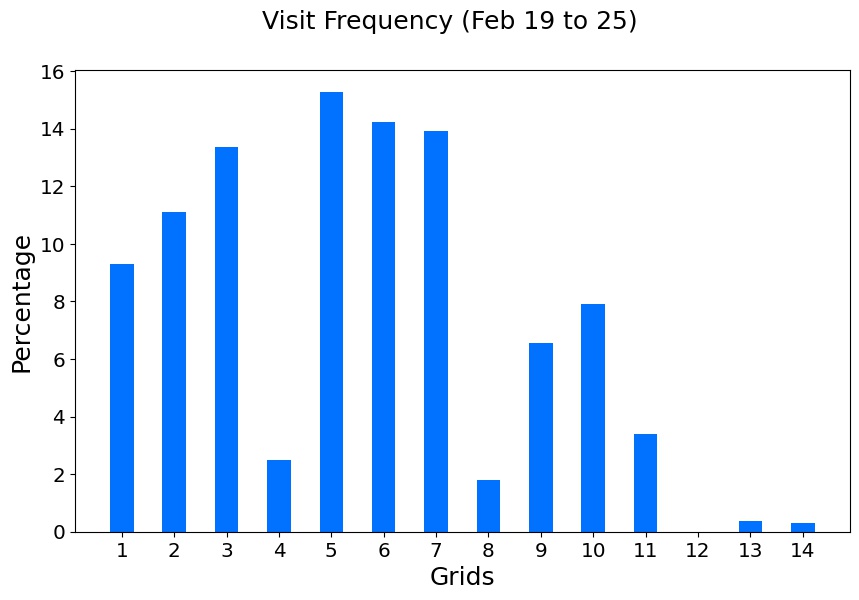}
    } \\
    \subfloat[Sutton Tools - occupancy rate
    \label{subfig:st-grid_occupy_portion_bar}]{
    \includegraphics[width=0.45\linewidth]{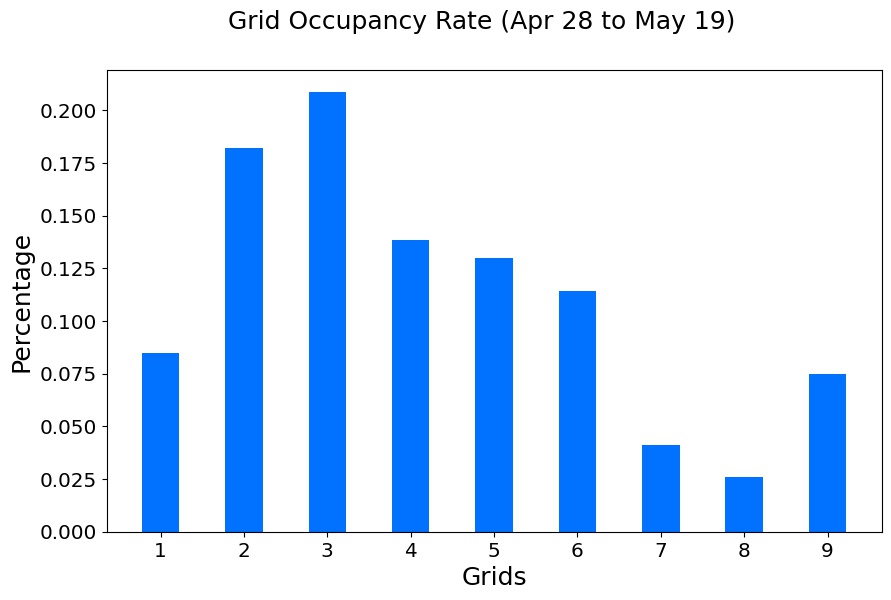}
    }
    \subfloat[Sutton Tools - visit frequency 
    \label{subfig:st-grid_visiting_freq_bar}]{
    \includegraphics[width=0.45\linewidth]{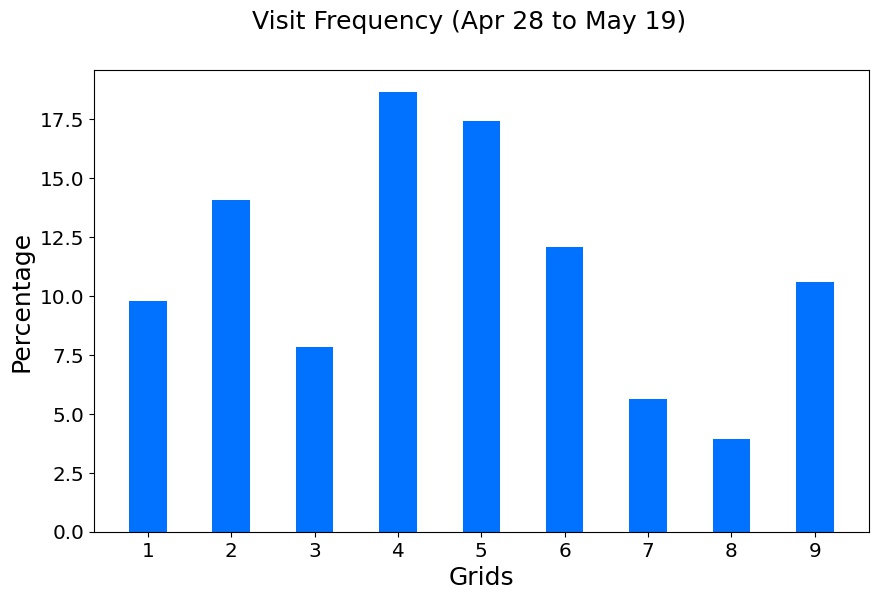}
    }
  \caption{The figures of occupancy rate and visit frequency by grids on two study locations.}
  \label{fig:occupancy_rate_visit_freq_bar}
\end{figure}

The heat maps and the corresponding bar charts of occupancy rates by grids can indicate the hotspot areas where the participants prefer to dwell. From Fig. \ref{subfig:mem_grid_occupy_portion} and \ref{subfig:memko-grid_occupy_portion_bar}, we can see that grids 7 and 3 of Memko have higher dwelling times than the other zones. Some tables for dining are located in grid 7, and most of the participants stay around the seats in grids 3 and 7. In Sutton Tools, Fig. \ref{subfig:st_grid_occupy_portion} and \ref{subfig:st-grid_occupy_portion_bar} show that the occupancy rate is more considerable in grids 2 and 3, which have the central workbench and a console for workers to conduct some routine tasks, respectively.

\subsubsection{Visit Frequency}
The Figs. \ref{subfig:mem_grid_visit_freq} and \ref{subfig:st_grid_visit_freq} reveal the zones visited frequently by participants in two different sites. And the corresponding figures of visit frequency are given in Figs. \ref{subfig:memko-grid_visiting_freq_bar} and Figs. \ref{subfig:st-grid_visiting_freq_bar}. In Memko, we found that more visits happened around the areas of the corridor, including grids 5, 6, and 7. More visits are observed in grid four and grid 5 of the Sutton Tools workshop, as these two zones place most of the components and the products of orders.

\subsubsection{Correlation Analysis}
\begin{figure*}[ht]
  \centering
    \subfloat[Memko grid 2
    \label{subfig:Memko_grid2_visit_air}]{
    \includegraphics[width=0.23\linewidth]{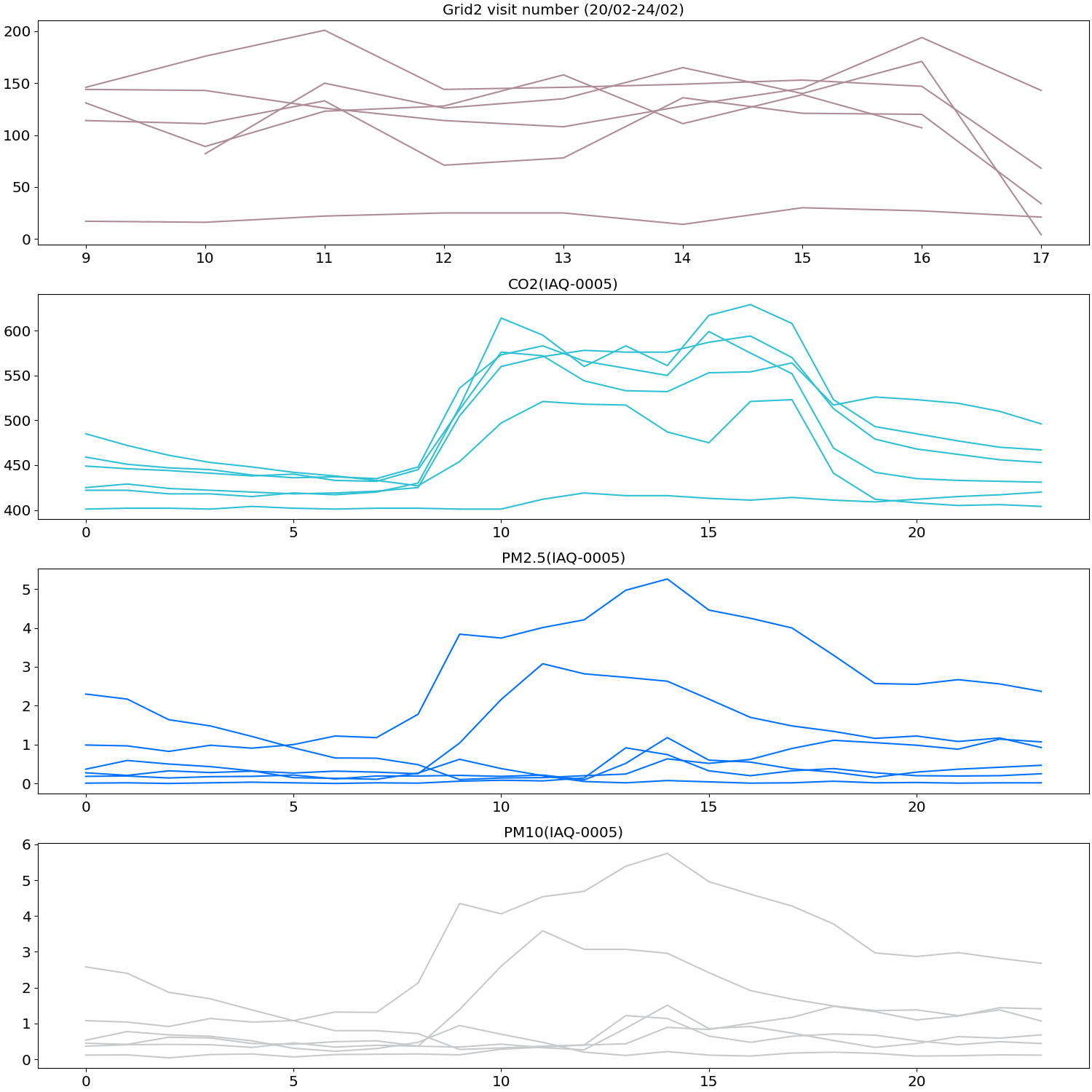}
    } \hspace{0.02cm}
    \subfloat[Memko grid 3
    \label{subfig:Memko_grid3_visit_air}]{
    \includegraphics[width=0.23\linewidth]{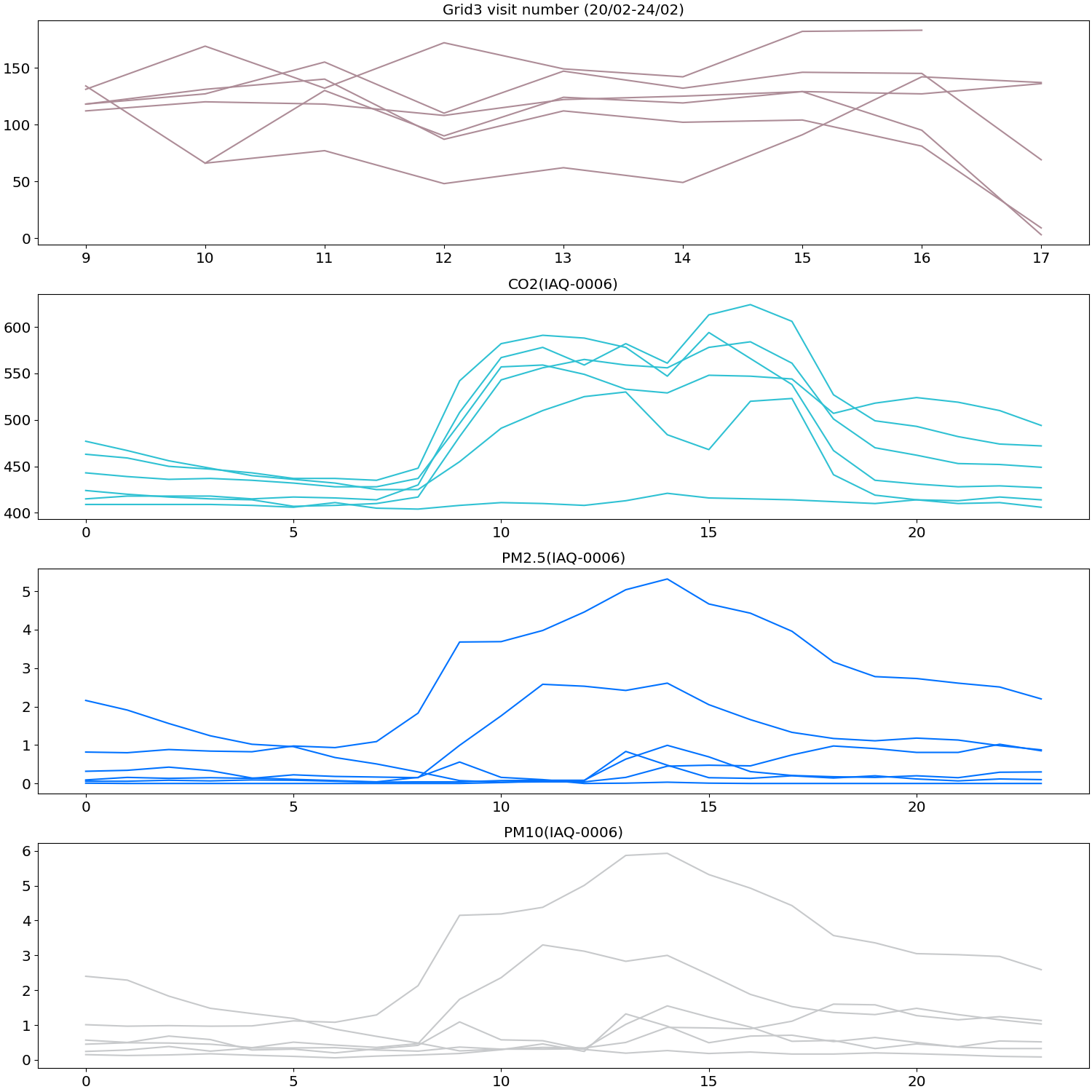}
    } \hspace{0.02cm}
    \subfloat[Memko grid 7
    \label{subfig:Memko_grid7_visit_air}]{
    \includegraphics[width=0.23\linewidth]{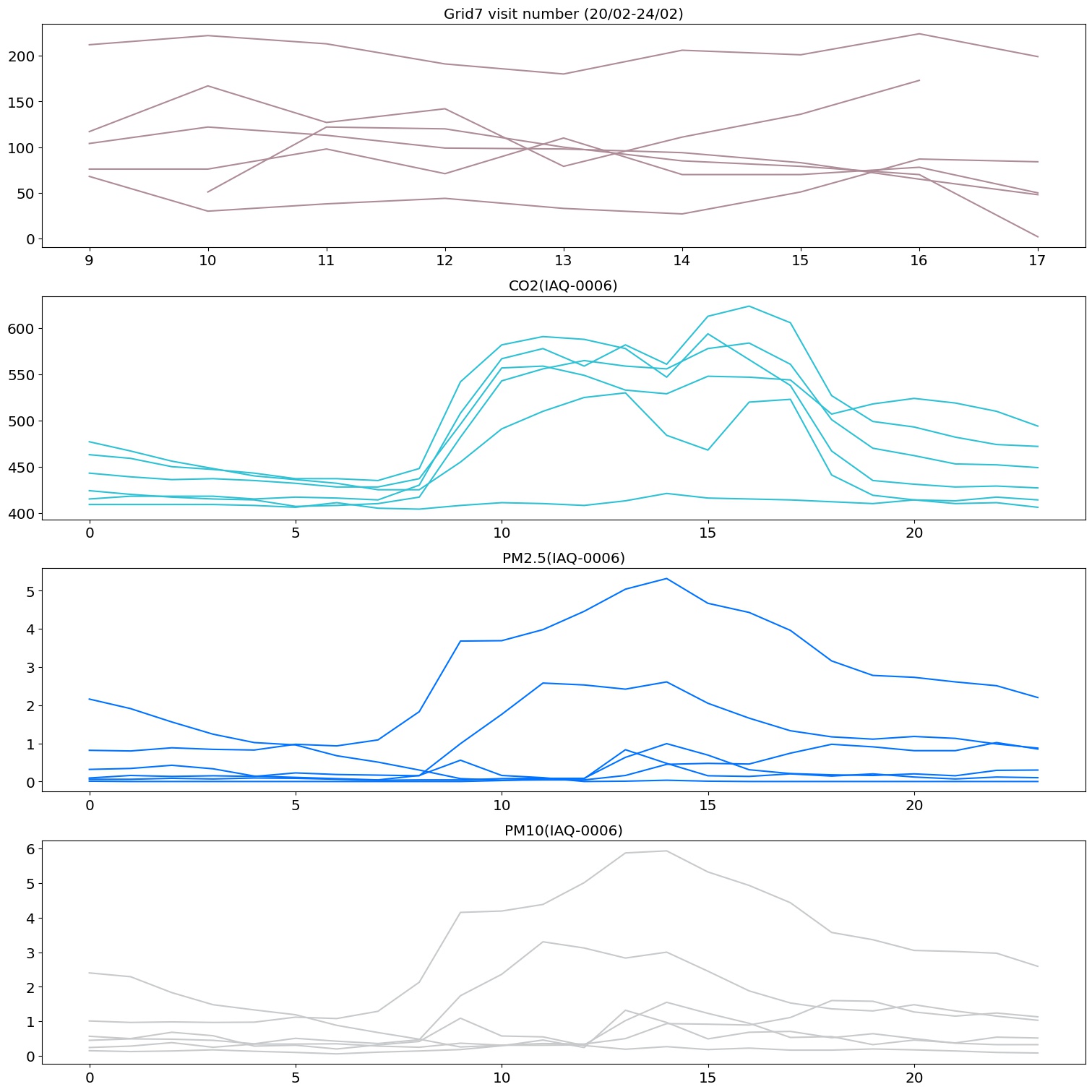}
    } \hspace{0.02cm}
    \subfloat[Memko grid 10
    \label{subfig:Memko_grid10_visit_air}]{
    \includegraphics[width=0.23\linewidth]{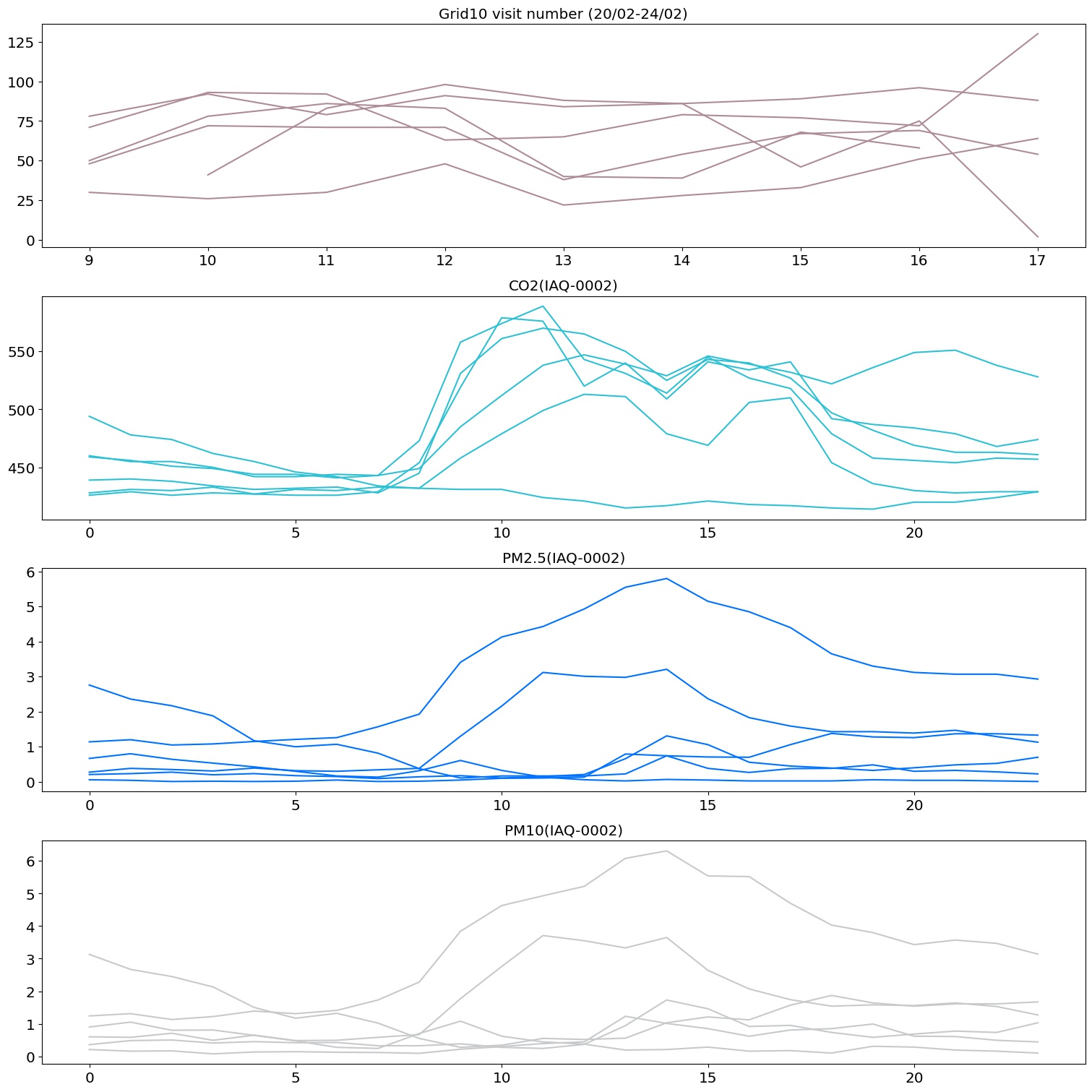}
    } \hspace{0.02cm}
    \subfloat[ST grid 1
    \label{subfig:st_grid1_visit_air}]{
    \includegraphics[width=0.23\linewidth]{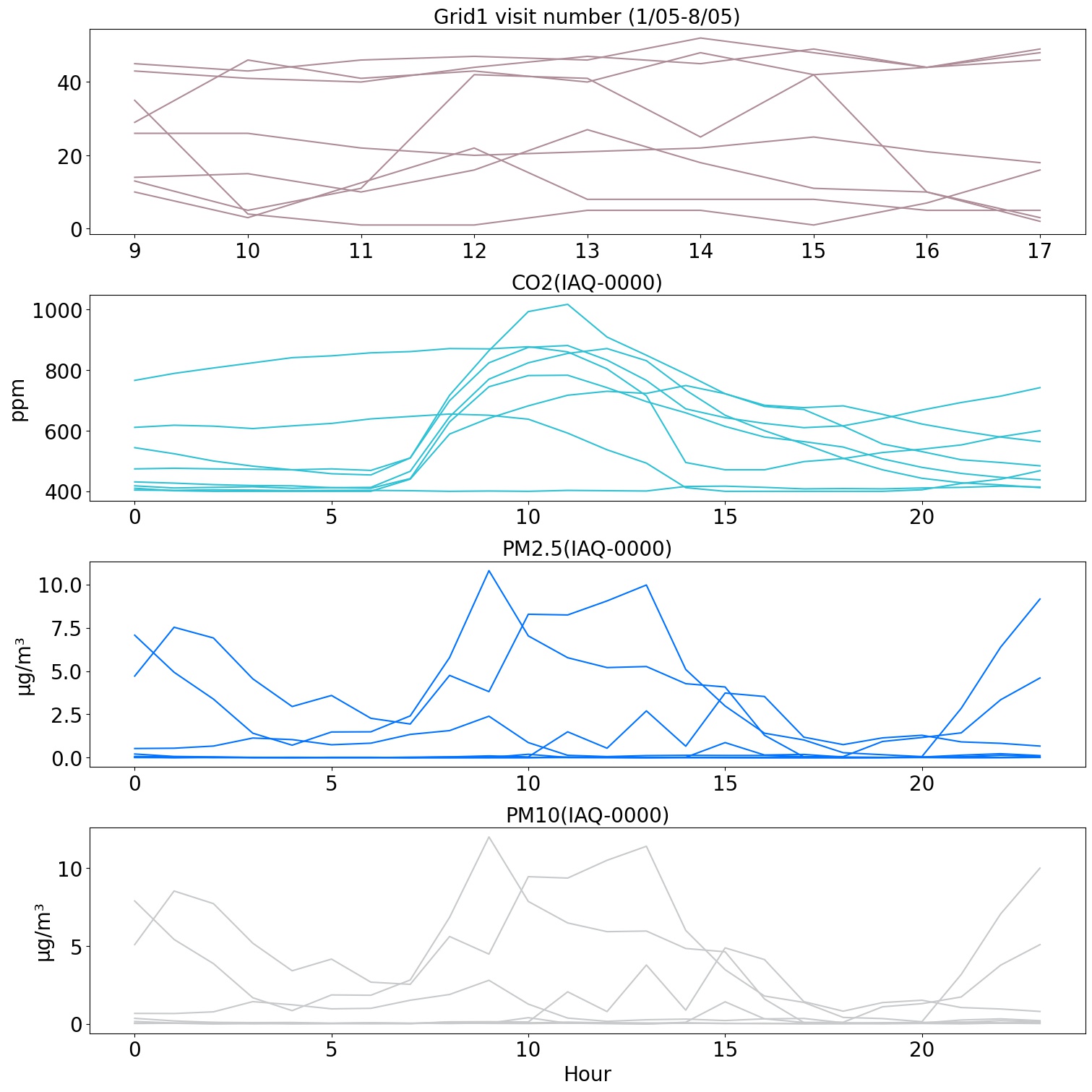}
    } \hspace{0.02cm}
    \subfloat[ST grid 2
    \label{subfig:st_grid2_visit_air}]{
    \includegraphics[width=0.23\linewidth]{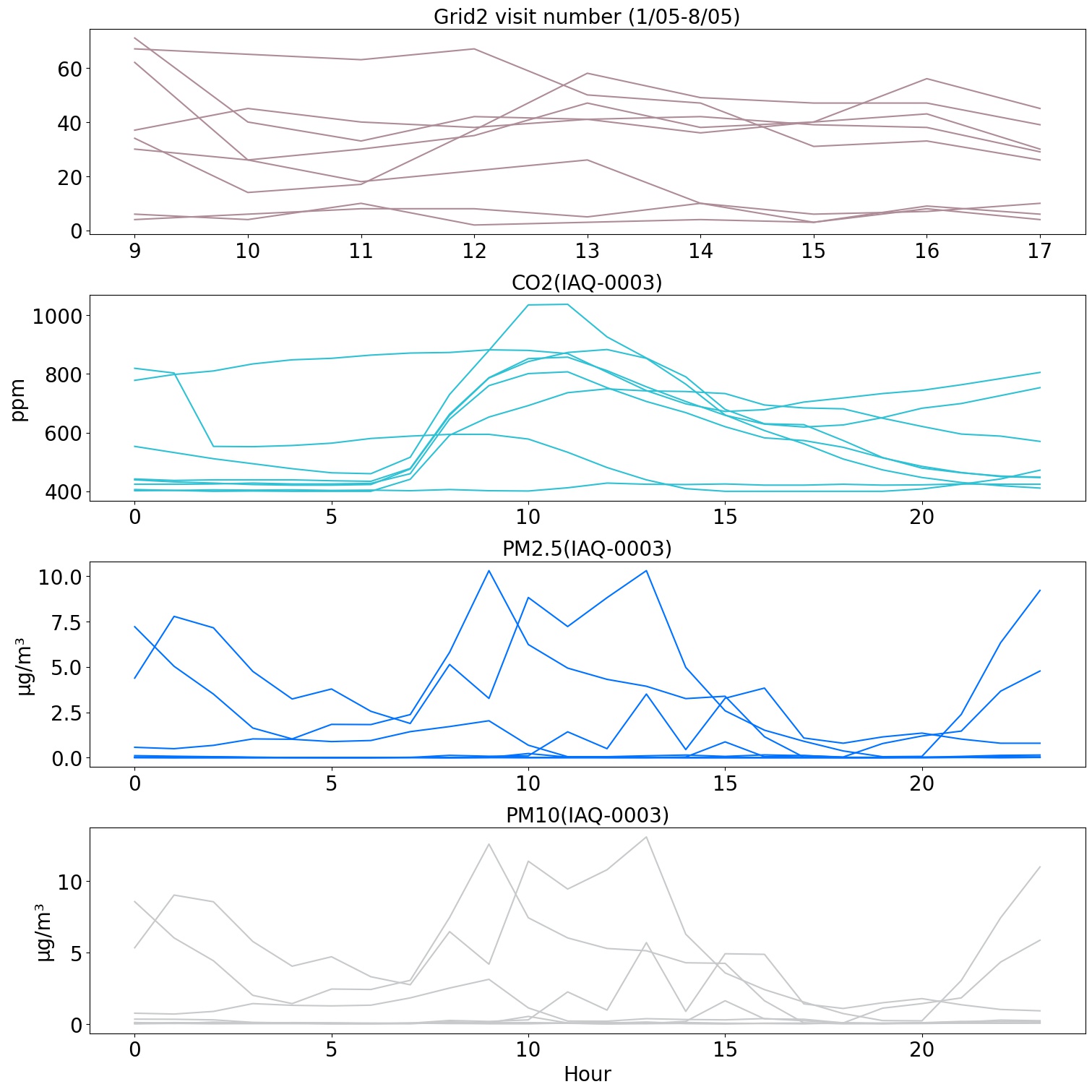}
    } \hspace{0.02cm}
    \subfloat[ST grid 3
    \label{subfig:st_grid3_visit_air}]{
    \includegraphics[width=0.23\linewidth]{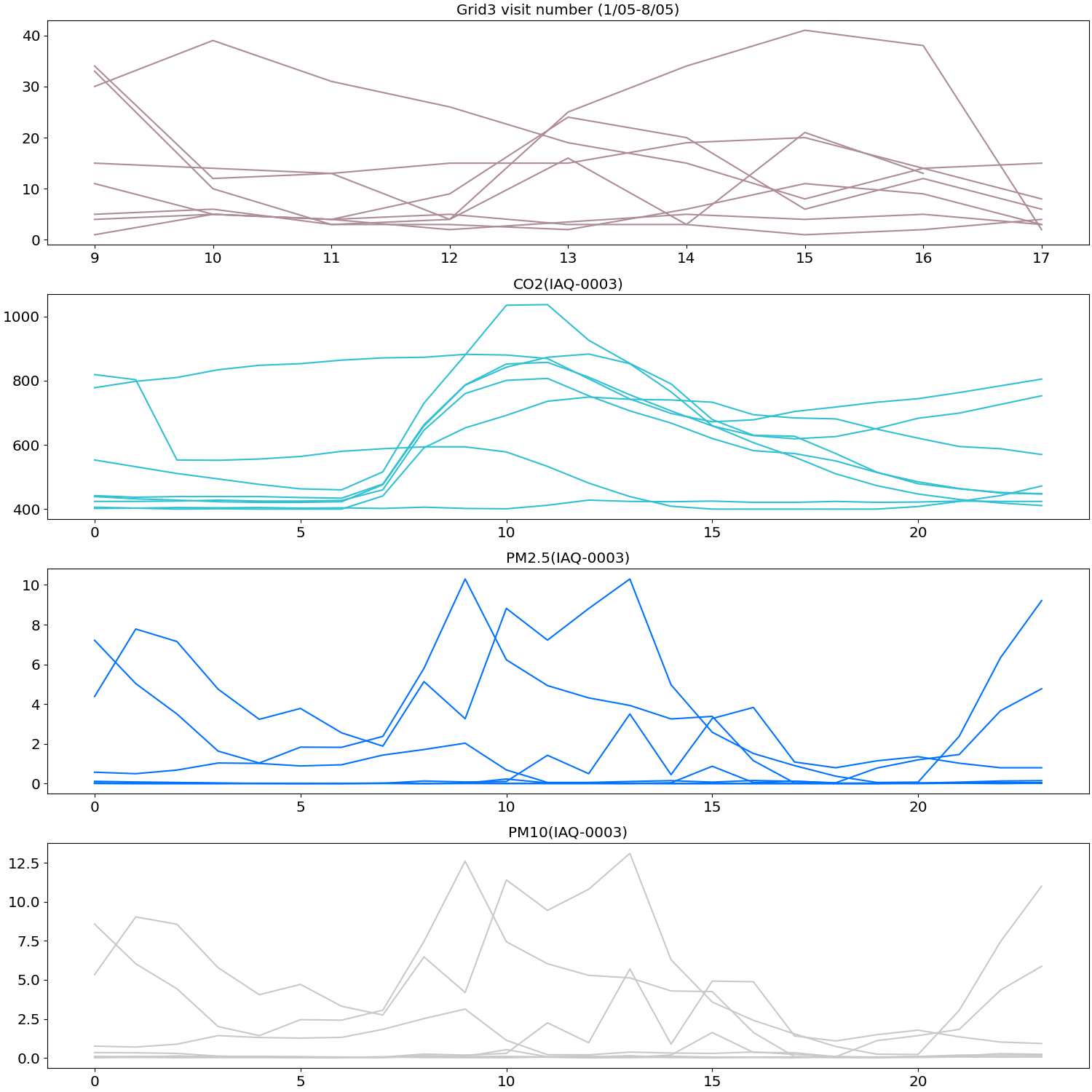}
    } \hspace{0.02cm}
    \subfloat[ST grid 5 
    \label{subfig:st_grid5_visit_air}]{
    \includegraphics[width=0.23\linewidth]{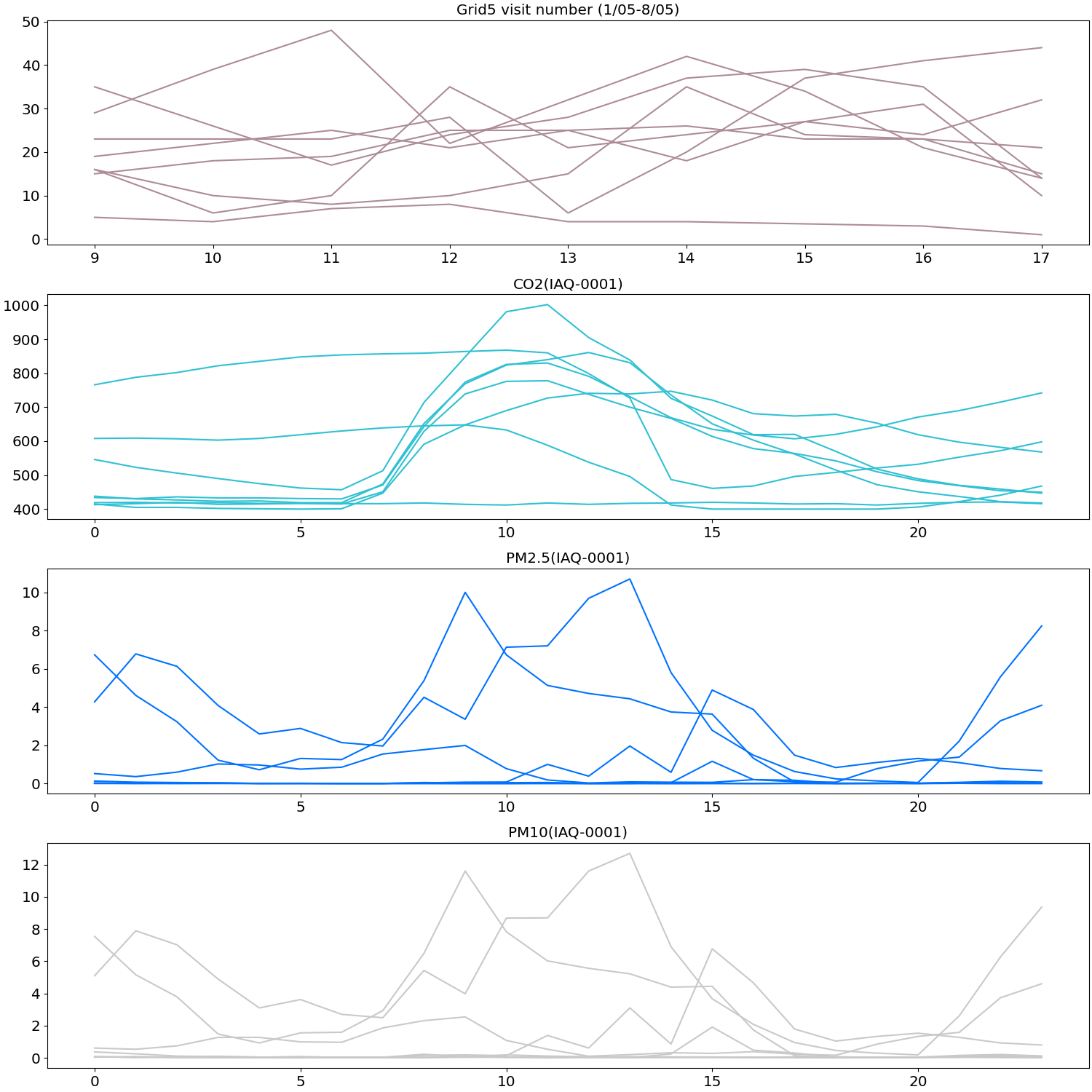}
    } 
  \caption{Visit number VS air quality in the grids of two study areas.}
  \label{fig:visit_number_air}
\end{figure*}

Ultimately, we demonstrate the changes in air quality with human visit statistics in Fig. \ref{fig:visit_number_air}. Each diagram represents the figure of visit number on one grid with the installation of the nearest indoor air quality sensor, and the deployment of devices is given in Fig. \ref{fig:sensors_deploy}. As far as we know, working hours are around 8 AM to 6 PM in both places. The results show that CO$_2$, PM2.5, and PM10 are relatively higher during busy hours and drop before and after working time; the pattern of change is consistent in observing all the main grids. We could claim that there is a significant influence on indoor air quality when the indoor visitor numbers have shifted.

\section{Conclusion}
\label{sec:conclude}
The global COVID-19 pandemic revealed that people are vulnerable to virus infections by airborne transport in indoor environments. Studying the association between indoor air quality and human movements could offer valuable insights to help control the risk of virus transmission. Motivated by the availability of various sensor devices for recording indoor climate conditions and tracking indoor subjects, we developed an integrated system to monitor indoor air quality and human movements simultaneously. The sensor networks detect and generate real-time data stored in the cloud for statistical analysis and particular mining tasks. We inferred the trajectories of anonymous participants based on carried BLE beacon signals. Then, the study reveals the correlation between human mobility patterns and IAQ features. The experimental results show a practical impact on the changes in air quality by human movements during different working hours. In the future, we intend to apply more recent techniques for building indoor human tracking systems, such as sensor deployment optimization or using more advanced unsupervised learning methods for human trajectory modeling. In real-life scenarios such as shopping malls where individuals move without wearing Bluetooth beacons, we can apply dead reckoning based on mobile sensors of users or triangulation based on the WiFi connection approach of mobile devices. In addition, we can extend the analysis of the association between IAQ and human mobility by implementing effective machine-learning models to forecast the trend of IAQ changes based on movement patterns.

\section*{Acknowledgment}

This research was funded by the Victorian Higher Education State Investment Fund—Project Digital and Business Transformation—Digitization for Safe Workplaces, Grant Number: VHESIF 3.2, Project ID PRJ00000201. We thank our industry partners (Memko Pty Ltd. and Sutton Tools Pty Ltd.) for their in-kind support and participation in the experimental studies. In addition, this project is also partially funded by the ARC Centre of Excellence for Automated Decision-Making and Society (CE200100005).

\bibliographystyle{IEEEtran}
\bibliography{main}


\end{document}